\documentclass[aps,prl,twocolumn,showpacs,preprintnumbers,superscriptaddress,
amsmath,amssymb]{revtex4}
\usepackage{CJK}
\usepackage{graphicx}% Include figure files
\usepackage{dcolumn}% Align table columns on decimal point
\usepackage{bm}% bold math

\begin{document}
\begin{CJK*}{UTF8}{gkai}
\title{Phase Diagram of Colloidal Hard Superballs: from Cubes via Spheres to Octahedra}

\author{Ran Ni (倪冉)} \email{r.ni@uu.nl}
\affiliation{Soft Condensed Matter, Utrecht University, Princetonplein 5, 3584 CC Utrecht, The Netherlands}
\author{Anjan Prasad Gantapara}
\affiliation{Soft Condensed Matter, Utrecht University, Princetonplein 5, 3584 CC Utrecht, The Netherlands}
\author{Joost de Graaf}
\affiliation{Soft Condensed Matter, Utrecht University, Princetonplein 5, 3584 CC Utrecht, The Netherlands}
\author{Ren\'{e} van Roij}
\affiliation{Institute for Theoretical Physics, Utrecht University, Leuvenlaan 4, 3504 CE Utrecht, The Netherlands}
\author{Marjolein Dijkstra} \email{m.dijkstra1@uu.nl}
\affiliation{Soft Condensed Matter, Utrecht University, Princetonplein 5, 3584 CC Utrecht, The Netherlands}

\date{\today}% It is always \today, today,

\begin{abstract}
The phase diagram of colloidal hard superballs, of which the shape interpolates between cubes and octahedra via spheres, is determined by
free-energy calculations in Monte Carlo simulations. We discover not only a stable face-centered cubic (fcc) plastic crystal phase for
near-spherical particles, but also a stable body-centered cubic (bcc) plastic crystal close to the octahedron shape. Moreover, coexistence of these two plastic crystals is observed with a substantial density gap. The plastic fcc and bcc crystals are, however, both unstable in the cube
and octahedron limit, suggesting that the rounded corners of superballs play an important role in stablizing the rotator phases. In addition, we
observe a two-step melting phenomenon for hard octahedra, in which the Minkowski crystal melts into a metastable bcc plastic crystal before
melting into the fluid phase.
\end{abstract}


\maketitle
\end{CJK*}
%\section{Introduction}
Recent breakthroughs in particle
synthesis have resulted in a spectacular variety of
anisotropic nanoparticles such as cubes, octapods, tetrapods, octahedra,
icecones, etc.~\cite{glotzer2007}.
A natural starting point to study the self-assembled structures
of these colloidal building blocks is to view them as hard particles [1]. Not
only can these hard-particle models be used to  predict properties of suitable
experimental systems,
but such models also provide a stepping stone towards
systems where soft interactions play a role~\cite{yethiraj2003,nanocube2011}.
Moreover, the analysis of hard particles is of
fundamental relevance
and raises problems that influence fields as diverse
as (soft) condensed matter~\cite{hynninen2007,glotzer2009,glotzer2007,escobedo},
mathematics~\cite{glotzer2009,torquatonature}, and
computer science~\cite{spherepack}.
In this light the concurrent boom in simulation studies of hard
anisotropic particles is not
surprising~\cite{glotzer2009,escobedo,torquatonature,joost,jiao2009,
batten2010,marechal2011,frank,vega1997,marechal2008,peter1997}.

The best-known hard-particle system consists of hard spheres, which freeze into
close-packed
hexagonal (cph) crystal structures~\cite{spherepack}, of which the ABC-stacked
cph crystal, better known as the face-centered
cubic (fcc) crystal phase, is thermodynamically stable~\cite{peterfcc1997}. Hard anisotropic
particles can form liquid-crystalline equilibrium states if they are
sufficiently rod- or disclike~\cite{marechal2011,peter1997}, but particles with shapes that are
close-to-spherical tend to order into plastic
crystal phases, also known as rotator phases~\cite{vega1997,marechal2008,peter1997}. In fact, simple
guidelines were recently proposed to predict the plastic- and liquid-crystal
formation only on the basis of rotational symmetry and shape anisotropy of hard
polyhedra~\cite{escobedo}. In this Letter we will take a different approach, based on
free-energy calculations,
 and address the question whether and to what extent rounding
the corners and faces of polyhedral particles
affects the phase behavior. Such curvature effects are of direct
relevance to experimental systems, in which sterically and charged stabilised
particles can often {\em not}
be considered as perfectly flat-faced and sharp-edged~\cite{rossi2011}. For instance,
recent experiments on nanocube assemblies show a continuous
phase transformation between simple cubic and
rhombohedral phases by increasing the ligand thickness
and hence the particle sphericity~\cite{nanocube2011}.

In this Letter, we study a system of colloidal hard superballs
in order to address these problems. A superball
is defined by the inequality
\begin{equation}
|x|^{2q} + |y|^{2q} + |z|^{2q} \le 1,
\end{equation}
where $x$, $y$ and $z$ are scaled Cartesian coordinates with $q$ the
deformation parameter. The shape of the superball interpolates
smoothly between two Platonic solids, namely
the octahedron ($q = 0.5$) and the cube ($q = \infty$) via the
sphere ($q = 1$). By determining the phase diagram of
these superballs as a function of $q$, we discovered a thermodynamically
stable body-centered cubic (bcc) plastic crystal
phase for octahedron-like superballs. To the best of our knowledge {\em no} plastic crystals
other than cph structures have so far been observed for hard particles.
Moreover, we find
that bcc and fcc plastic crystal phases are unstable
for hard octahedra and hard cubes, respectively. Therefore,
rounded faces and edges may play an important role in
stabilizing rotator phases, while flat faces tend to stabilize crystals.

\begin{figure}
\includegraphics[width=0.5\textwidth]{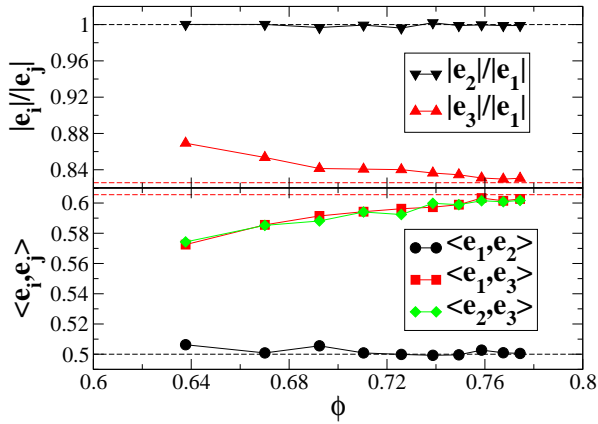}
\caption{\label{fig1}(Color online) The deformation of the crystal unit cell with
lattice vectors $\mathbf{e}_{i}$ as a function of
packing fraction $\phi$ in a system of hard superballs with $q=2.5$. The
dashed lines in the figures indicate the values for the $C_1$ crystal.}
\end{figure}

\begin{figure*}
\includegraphics[width=0.85\textwidth]{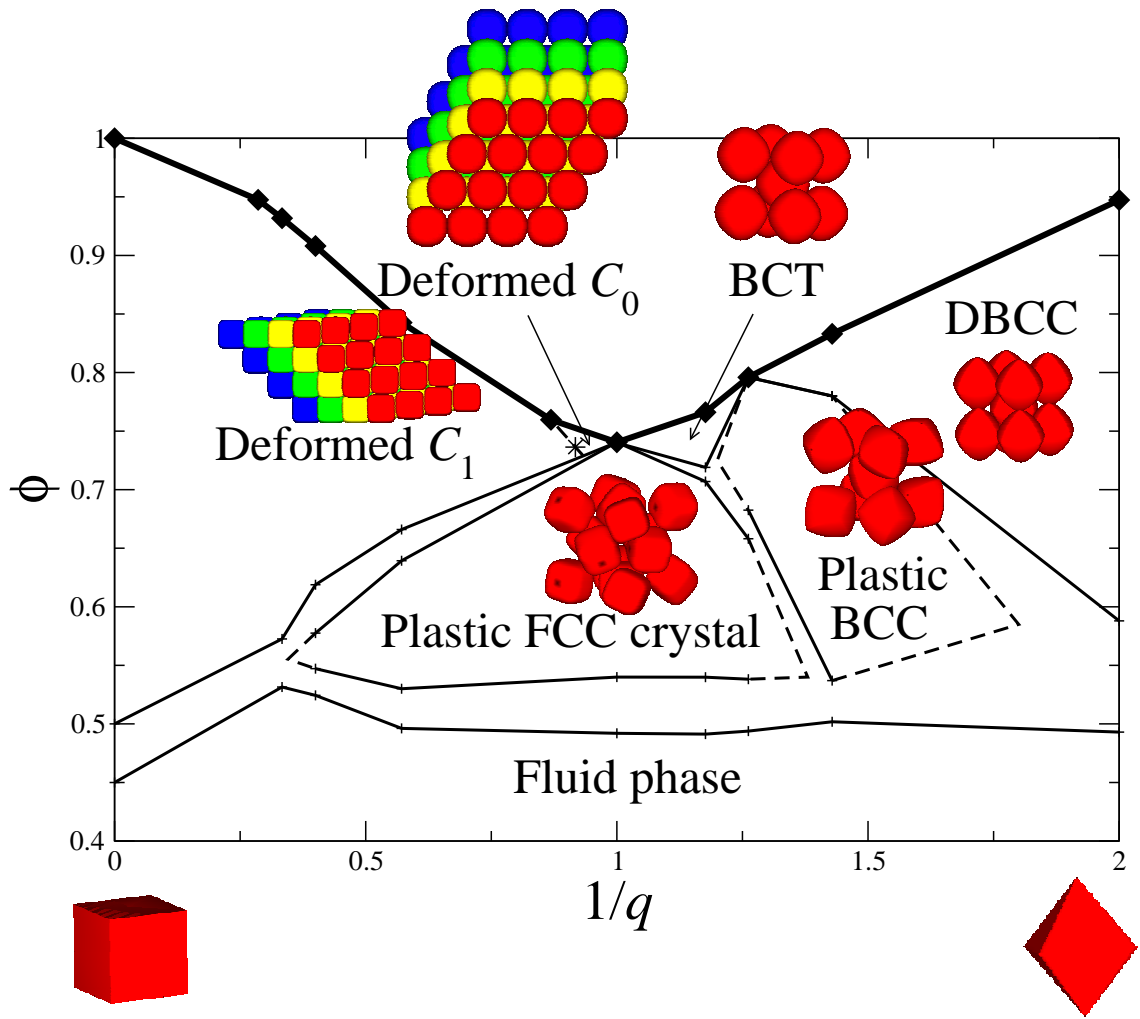}
\caption{\label{phasediag}(Color online) Phase diagram for hard superballs
in the $\phi$ (packing fraction) versus $1/q$ representation where $q$ is the
deformation parameter. Here the $C_1$ and $C_0$ crystals are defined in
Ref.~\cite{jiao2009}, where the particles of the same color are in the same layer of stacking.
 The solid diamonds indicate the close packing, and the
locations of triple points are determined by extrapolation as shown by the dashed
lines. The phase boundaries for hard cubes are taken from Ref.~\cite{frank}.}
\end{figure*}

Following Refs.~\cite{filion2009,joost}, we first calculate the
close-packed structures for systems of hard superballs. We employ the algorithm of checking overlap described in Ref.~\cite{supinfo}. For cube-like particles,
it is found that at close packing there are so-called $C_0$ and $C_1$ crystal
phases in accordance with Ref.~\cite{jiao2009}. When we
perform $NPT$ Monte Carlo simulations with variable box shape to determine
the equation of state (EOS) of the crystal phase, our simulation results show
that both the $C_0$ and the $C_1$ crystals deform with decreasing density.
 The lattice vectors for $C_1$ crystals are given by $\mathbf{e}_1 =
2^{1-1/2q}\,\mathbf{i}+2^{1-1/2q}\,\mathbf{j}$
,$\mathbf{e}_2=2^{1-1/2q}\,\mathbf{i}+2^{1-1/2q}\,\mathbf{k}$ ,
$\mathbf{e}_3=2(s+2^{-1/2q})\,\mathbf{i}-2s\,\mathbf{j}-2s\,\mathbf{k}$,
where $\mathbf{i}$, $\mathbf{j}$ and $\mathbf{k}$ are the unit vectors
along the axes of the particle, $s$ is the smallest positive root of the
equation $(s+2^{-1/2q})^{2q}+2s^{2q}-1=0$, and there is one particle in the unit
cell~\cite{jiao2009}.
For instance, when $q=2.5$, one finds that
$\langle\mathbf{e}_1,\mathbf{e}_2\rangle = 0.5$,
$\langle\mathbf{e}_1,\mathbf{e}_3\rangle =
\langle\mathbf{e}_3,\mathbf{e}_2\rangle = 0.60552$,
$\left|\mathbf{e}_2\right|/\left|\mathbf{e}_1\right| = 1$ and $
\left|\mathbf{e}_3\right|/\left|\mathbf{e}_1\right| = 0.825737$,
where $\langle\mathbf{e}_i,\mathbf{e}_j\rangle$ is the cosine of
angle between $\mathbf{e}_i$ and $\mathbf{e}_j$. The calculated angles and the
length ratios between lattice vectors as a function of packing fraction $\phi$
for the
cube-like particles with $q=2.5$ are shown in Fig.~\ref{fig1}. We find that
at packing fractions approaching close packing, the crystal remains in the $C_1$
phase. With decreasing packing fraction, the crystal lattice deforms towards an
fcc structure: $\langle\mathbf{e}_1,\mathbf{e}_2\rangle
= \langle\mathbf{e}_1,\mathbf{e}_3\rangle =
\langle\mathbf{e}_2,\mathbf{e}_3\rangle= 0.5$ and $
\left|\mathbf{e}_2\right|/\left|\mathbf{e}_1\right|  =
\left|\mathbf{e}_3\right|/\left|\mathbf{e}_1\right| = 1$.
Moreover, when $1 < q < 3$, it is found that the deformed $C_0$ and deformed $C_1$ crystal melt into an fcc plastic crystal phase. By Einstein
integration, we calculated the Helmholtz free energy as a function of packing fraction for both the fcc plastic crystal and the deformed
$C_1$/$C_0$ crystal phases~\cite{frenkelbook}. Combined with the free-energy calculations for the fluid phase done by Widom's particle insertion
method, we obtain the phase boundaries in the phase diagram shown in Fig.~\ref{phasediag}. The part of the phase diagram for hard cube-like
superballs roughly agrees with the empirical phase diagram by Batten {\em et al.}~\cite{batten2010} At high packing fractions, there are stable
deformed $C_0$ and $C_1$ phases. When $q > 1.1509$, the close-packed structure is the $C_1$ crystal, whereas it is the $C_0$ crystal whenever
$1<q<1.1509$~\cite{jiao2009}. To determine the location of the transition from the deformed $C_0$ crystal to the deformed $C_1$ crystal, we
performed two series of $NPT$ MC simulations with increasing value of $q$ for the first series and decreasing $q$ for the second series of
simulations at pressure $P^*=Pv/k_BT \simeq 250$, with $k_B$ the Boltzmann constant, $T$ the temperature, and $v$ the volume of the
particle~\cite{marechal2011}. The first series started from a $C_0$ crystal phase, while the second series of simulations started from a $C_1$
crystal phase. Our simulations show that the phase transition occurred around $q=1.09$ at packing fraction $\phi = 0.736$ as shown by the asterix
in Fig.~\ref{phasediag}. Moreover, for hard cubes ($q=\infty$) the $C_1$ crystal is a simple cubic (sc) crystal. Although it was found that for hard cubes there is a significant amount of vacancies in the simple cubic crystal, it only shifts the phase boundary by $\sim 2\%$ in packing fraction~\cite{frank}. In our simulations, we did not observe any vacancies in the crystals of hard superballs with $q \le 3$, we therefore assume that the possible presence of vacancies would not shift the phase boundary significantly.

The other part of the phase diagram concerns the octahedron-like superballs.
For $0.79248<q<1$, we obtained a denser structure than the predicted $O_0$
lattice of Ref.~\cite{jiao2009}. For instance, after compressing
the system to pressures around $P^*=10^7$ at $q=0.85$, we obtained a body-centered-tetragonal (bct) crystal
with $\phi=0.7661$. This is denser than the $O_0$ crystal, which
achieves $\phi=0.7656$ at $q=0.85$. Note however that these two crystals are very
similar to each other, since $O_0$ is also a form of a bct lattice. The only
difference is that the orientation of the particles in the $O_0$ crystal
is the same as the symmetry of the axes in the crystal lattice, while in our bct
crystal there is a small angle between these two orientations in the square
plane of the crystal.
Furthermore, for $q<0.79248$, we also found a crystal with denser packing than the predicted $O_1$ crystal in Refs.~\cite{jiao2009}. For $q=0.7$, we performed floppy-box MC simulations with
several particles to compress the system to a high pressure state, i.e.,
$P^*=10^7$. We found a deformed bcc (dbcc) crystal shown in
Fig.~\ref{phasediag}, which is an intermediate form between the bcc lattice and
the Minkowski crystal~\cite{mink}. The lattice vectors are
$\mathbf{e}_1 = 0.912909 \mathbf{i} + 0.912403\mathbf{j}-0.912165\mathbf{k}$,
$\mathbf{e}_2 = -0.271668 \mathbf{i} + 1.80916\mathbf{j}-0.288051\mathbf{k}$ and
$\mathbf{e}_3 = 0.28834 \mathbf{i} -0.272001\mathbf{j} -1.80882\mathbf{k}$, where $\mathbf{i}$, $\mathbf{j}$ and $\mathbf{k}$ are the unit vectors along the axes of the particle.
Our dbcc crystal is very close the predicted $O_1$ crystal, whose lattice vectors are
$\mathbf{e}_1 = 0.912492\mathbf{i}+0.912492\mathbf{j}-0.912492\mathbf{k}$,
$\mathbf{e}_2 = -0.2884 \mathbf{i}+1.80629\mathbf{j} -0.2884\mathbf{k}$, and
$\mathbf{e}_3 = 0.2884 \mathbf{i}-0.2884 \mathbf{j} -1.80629 \mathbf{k}$. However, it has a packing fraction of $\phi = 0.832839$ which is denser than the predicted $O_1$ crystal with $\phi = 0.824976$ in Refs.~\cite{jiao2009} by roughly $1\%$.
In Ref.~\cite{jiao2009}, the $O_0$ and $O_1$ phases are
found to switch at $q=0.79248$. We also observed that the bct and dbcc crystals
both transform into the bcc phase at $q=0.79248$.

\begin{figure}[h]
\includegraphics[width=0.5\textwidth]{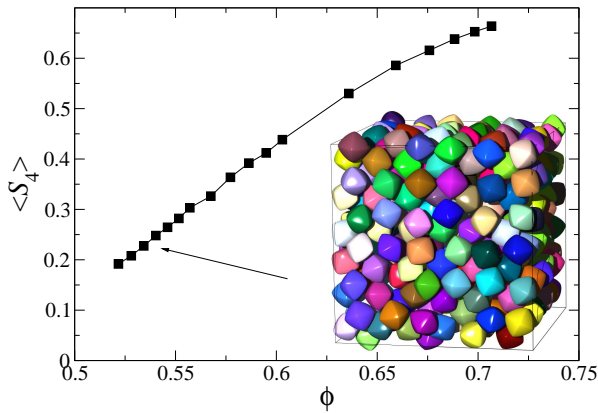}
\caption{\label{fig2}(Color online) Cubatic order parameter $S_4$ as a function of
packing fraction $\phi$ for a bcc plastic crystal phase of hard superballs with $q=0.7$. The inset shows a typical
configuration of a bcc plastic crystal of hard superballs with $q=0.7$ at $\phi=0.54$.}
\end{figure}

As shown in Fig.~\ref{phasediag}, when the shape of the superballs is close to spherical, i.e., $0.7<q<3$, there is always a stable fcc plastic
crystal phase. Surprisingly, when the shape of superballs is octahedron-like, we find a stable bcc plastic crystal phase. Moreover, around
$q=0.8$ we even find a fairly broad two-phase regime where a low-density fcc plastic crystal coexists with a high-density bcc plastic  crystal phase.
In order to quantify the orientational order in the bcc plastic crystal, we calculate the cubatic order parameter $S_4$ given
by~\cite{batten2010}
\begin{equation}
 S_4 =
\max_{\mathbf{n}}{\left\{\frac{1}{14N}\sum_{i,j}{\left(35|\mathbf{u}_{ij}
\cdot \mathbf{n} |^4 - 30|\mathbf{u}_{ij} \cdot \mathbf{n} |^2 +3
\right)}\right\}},
\end{equation}
where $\mathbf{u}_{ij}$ is the unit vector of the $j$-th axis of particle $i$,
$N$ the number of particles, and ${\bf n}$ is a unit vector. The cubatic order parameter $S_4$ is shown in Fig.~\ref{fig2} as a
function of packing fraction for a  system of superballs with $q=0.7$.
We observe that $\langle
S_4\rangle \simeq 0.2$ at low packing fractions, which means that there is a very weak orientational
order in the system~\cite{supinfo}. With increasing packing fraction, the cubatic order
parameter increases monotonically to around $0.65$ at a packing fraction of
$0.7$, which is indicative of a medium-ranged orientationally ordered
system.
This suggests that the entropic repulsion due to the rotation of the
octahedron-like superballs stablizes the bcc lattice.

\begin{figure}
\includegraphics[width=0.5\textwidth]{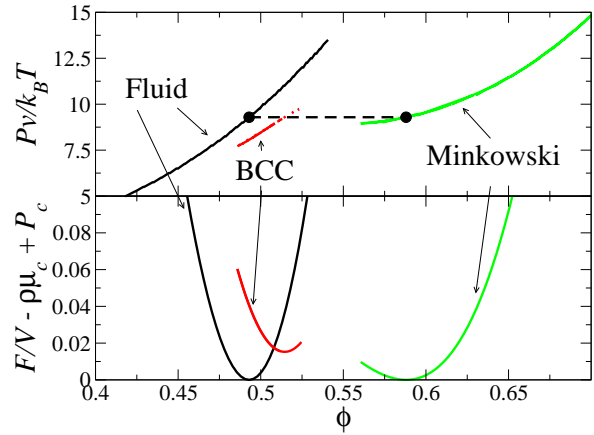}
\caption{\label{fig4}(Color online) A part of the equation of state for
hard octahedra. The pressure $Pv/k_BT$ and free-energy density
$F/V-\rho \mu_c + P_c$ as a function of packing fraction $\phi$. Here $v$ is
the volume of the particle; $F$ and $V$ are the Helmholtz free energy and the
volume of the system (in units of particle volume) respectively; $\mu_c$ and
$P_c$ are the chemical potential and pressure at bulk coexistence respectively with
$\rho$ the number density of the particles. The solid lines in the EOS for the Minkowski and
the bcc plastic crystal phases are obtained by melting the close-packed Minkowski crystal in
floppy box $NPT$ MC simulations, and the dotted line for the bcc plastic crystal is
obtained by compressing the crystal in cubic box $NPT$ MC simulations.
The black points and dashed line show the coexistence between the fluid phase and
the Minkowski crystal phase.}
\end{figure}

Due to the numerical  instability in the overlap algorithm, we are not able to
investigate systems of superballs with $q<0.7$~\cite{donev}. However, we can
use the separating axis theorem~\cite{escobedo} to simulate hard
superballs with $q=0.5$, i.e., perfect octahedra. When we compressed the system
from a fluid phase, we did not observe the spontaneous formation of a crystal
phase in our simulation box within our simulation time. When we expand
the Minkowski crystal, which is the close-packed
structure of octahedra, in $NPT$ MC simulations by decreasing the pressure, the
system melts into a bcc
plastic crystal phase as shown in Fig.~\ref{fig4}. We also calculated the free
energy for these three phases to determine the phase boundaries. To exclude
finite-size effects in the free-energy calculation of crystal phases, we
performed
Einstein integration for systems of $N=1024$, $1458$, and $2000$
particles, and applied a finite-size correction~\cite{frenkelbook}. We
confirmed the errors in the free-energy calculations to be on the
order of $10^{-3}k_BT$
per particle.
The calculated free-energy densities for the three phases are shown in
Fig.~\ref{fig4}. Employing a common tangent construction, we found that there is only
phase coexistence between a fluid phase and a Minkowski crystal phase, while the bcc
plastic crystal phase is metastable. However, the free-energy differences
between the fluid and the plastic crystal phase at the
bulk coexistence pressure is very small , i.e., $\sim
10^{-2}k_BT$ per
particle. Moreover, the Minkowski crystal melts into a bcc plastic crystal before melting
into the fluid phase.
Our results thus show that the rounded
corners of octahedra play an important role in stablizing the bcc plastic
crystal phase which is a new plastic crystal phase for systems of hard
particles.

In conclusion, using free-energy calculations we have determined the full phase diagram of hard superballs with shapes interpolating between cubes and
octahedra, i.e., $0.5 \le q < \infty $. In systems of cube-like superballs ($q>1$), we find a stable deformed $C_1$ phase at high packing fraction,
except close to the sphere-limit ($q=1$) where a deformed $C_0$ crystal is stable. For $q < 3$ the crystal phase melts into an fcc plastic crystal
before melting into a fluid phase of cubelike superballs. In systems of octahedron-like superballs ($0.5<q<1$), we find a stable bct or a deformed bcc crystal phase upon
approaching close packing, with a crossover at $q = 0.79248$. Moreover, a stable fcc plastic crystal appears at intermediate densities for $0.7 <
q \le 1$. Interestingly, for $q < 0.85$, we find a novel stable bcc plastic crystal phase, which can even coexist with the  fcc plastic crystal phase
at around $q=0.8$. More surprisingly, the bcc and fcc rotator phases are unstable for the flat-faced and sharp-edged hard octahedra and hard
cubes, respectively, which suggests that the rounded corners of superballs may play an important role in stabilizing rotator phases. It is tempting to argue that
entropic directional forces ~\cite{glotzer2011} that tend to align sufficiently large flat faces of  polyhedral-shaped particles destabilize rotator phases in
favor of crystals. We stress here that rounded corners are not a necessary condition for stable rotator phases since almost spherical polyhedral particles have been shown to form rotator phases as well \cite{escobedo}. Finally, we also observed a two-step melting phenomenon in the system of hard octahedra, such that the Minkowski crystal melts
into a metastable bcc plastic crystal before melting into the fluid phase. Nanoparticle self-assembly is therefore surprisingly sensitive to
particle curvature.

\begin{acknowledgments}
  We thank F. Smallenburg for fruitful
discussions and Dr. D. Ashton for making the snapshot in Fig.~\ref{fig2}.
We acknowledge the financial support from a NWO-VICI grant and
Utrecht University
High Potential Programme.
\end{acknowledgments}

%\bibliography{cubref}% Produces the bibliography via BibTeX.
%Merlin.mbs v4.21 2009-07-09.

\end{document}